\documentclass[journal=mamobx]{achemso}

\usepackage{graphicx}
\usepackage{amsmath,amssymb}
\usepackage{siunitx}
\usepackage{overpic}

\graphicspath{{figs/}}

\newcommand*{\vect}[1]{\mathbf{#1}}

\newcommand*{\lp}{\ell_{\rm P}}
\newcommand*{\we}{w_{\rm eff}}

\newcommand*{\fig}[1]{\ref{fig:#1}}
\newcommand*{\subfig}[2]{\ref{fig:#1}(#2)}
\newcommand*{\eq}[1]{\ref{eqn:#1}}

\newcommand*{\rv}{\vect{r}}
\newcommand*{\rn}{\vect{r}_0}
\newcommand*{\rp}{\vect{r}'}
\newcommand*{\rpr}{\vect{r}'}

\newcommand*{\pcn}{p_N^{(n)}}
\newcommand*{\pnn}{p_N^{(n)}}
\newcommand*{\rhonn}{\rho_N^{(n)}}

\author{E. Werner}
\affiliation{Department of Physics, University of Gothenburg, Sweden}
\author{F. Westerlund}
\affiliation{ Department of Chemical and Biological Engineering, Chalmers University of Technology, Sweden}
\author{J. O. Tegenfeldt}
\affiliation{Department of Physics, Division of Solid State Physics, Lund University, Sweden}
\author{B. Mehlig}
\email{bernhard.mehlig@physics.gu.se}
\affiliation{Department of Physics, University of Gothenburg, Sweden}

\title{Monomer distributions and intra-chain collisions of a polymer confined to a channel} 

\begin{document}

\date{\today}

\begin{abstract}
\noindent We study the conformations of a self-avoiding polymer confined to a channel by computing the cross-sectional distributions of the positions of its monomers. By means of Monte-Carlo simulations for a self-avoiding, freely-jointed chain we determine how the cross-sectional distribution for a given monomer depends on its location in the polymer, and how strongly this distribution is affected by self-avoidance. To this end we analyze how the frequency of intra-chain collisions between monomers depends on their spatial position in the channel and on their location within the polymer. We show that most collisions occur between closely neighboring monomers. 
As a consequence the collision probability depends only weakly on the spatial position of the monomers. Our results explain why the effect of self-avoidance on the monomer distributions is weaker than predicted by mean-field theory. We discuss the relevance of our results for studies of DNA conformations in nanofluidic channels.
\end{abstract}

\section{Introduction} 

\noindent The behavior of polymers in confined environments is of interest from both technological and fundamental perspectives. Nanofluidic channels have for example been used to stretch and visualize single DNA molecules\cite{tegenfeldt2004,reisner2005}. This is useful for mapping the sequence of intact long DNA molecules \cite{reisner2010,nyberg2012,lam2012}, as well as for monitoring protein-DNA interactions \cite{wang2005,riehn2005,persson2012}, and to explore fundamental polymer physics of DNA \cite{reisner2009,reisner2007,metzler2006,balducci2006}. Furthermore, polymers in living systems are generally found in crowded and confined environments, and the confinement influences both intra- and inter-polymer interactions\cite{zhou2008,marenduzzo2010,jun2006}. 
From a polymer physics perspective, confinement affects two key parameters of the DNA: its orientation and its density. The orientation can be monitored by polarization sensitive imaging \cite{persson2009} and can be used as a tool to understand the extension of the DNA\cite{werner2012}. Conversely the local density is expected to have a direct effect on how molecules access binding sites along DNA where higher density gives shorter diffusion times between DNA fragments\cite{li2009} but also greater steric hindrance leading to lower binding rates and for example impeded overall enzymatic activity\cite{bar2009,mcCalla2009}. Furthermore, for DNA condensation to occur, non-adjacent base pairs must approach each other\cite{baumann2000}.
To obtain a detailed understanding of molecular transport and molecular interactions within confined DNA it is therefore necessary to not only know the local concentration of confined DNA, but also how the distribution of a section of DNA depends on its position within the molecule. Unfortunately, these \textit{monomer distributions} are still difficult to determine experimentally, and analytical calculations and computer simulations are therefore important tools in this context. \\ \hspace*{\fill}

\noindent In the 1960s, it was realized that the monomer distributions of a confined polymer can be theoretically analyzed in terms of a diffusion-annihilation equation, with absorbing boundary conditions\cite{diMarzio1965}. The annihilation term represents the effect of self-avoidance, and is therefore absent for an ideal polymer. In this case the equation simplifies to a diffusion equation, which is easily solved for simple confining geometries, such as channels with circular, rectangular, 
or triangular \cite{dorfman2013} cross sections, or slits \cite{casassa1967}. Although the ideal model neglects intra-polymer interactions, it is frequently used to describe the monomer distribution of real polymers \cite{stein2006,freed2010,ramirez2010}. 
The reason is that calculating the monomer distributions of an interacting polymer is significantly more difficult. This is because the annihilation term in the diffusion-annihilation equation is determined by the collision frequency, which in turn is related to the monomer distributions\cite{diMarzio1965}. The conventional way of approximately closing this system of equations is called \textit{self-consistent field theory}\cite{deGennesBook}. It assumes that the collision probability can be explained by a mean-field theory, neglecting intra-chain correlations. \\ \hspace*{\fill}
                                                                                                
\noindent In this paper, we investigate the monomer distributions of a single polymer confined to a square channel with side length $D\gg b$, where $b$ is the \textit{Kuhn length} of the polymer\cite{rubinsteinBook}. To test the effect of self-avoidance upon the monomer distributions and collision frequency of a polymer confined to a channel, we performed Monte Carlo simulations for a polymer model where the excluded volume of each monomer can be varied. We find that mean-field theory underestimates the probability of intra-chain collisions, but overestimates the extent to which this probability depends on the position of a monomer in the channel. This theory thus overestimates the broadening effect of self-avoidance. 
We show that the collision probability can be qualitatively understood by modeling the polymer as a diffusing particle with a drift velocity in the channel direction\cite{werner2012}. Finally, we show that if the position dependent collision probability is known, the diffusion-annihilation equation correctly describes its impact on the monomer distributions of the self-avoiding polymer.

\section{Model and simulation results for the monomer distributions}
\begin{figure}[t]
 \centering
\begin{overpic}[width=3.25 in]{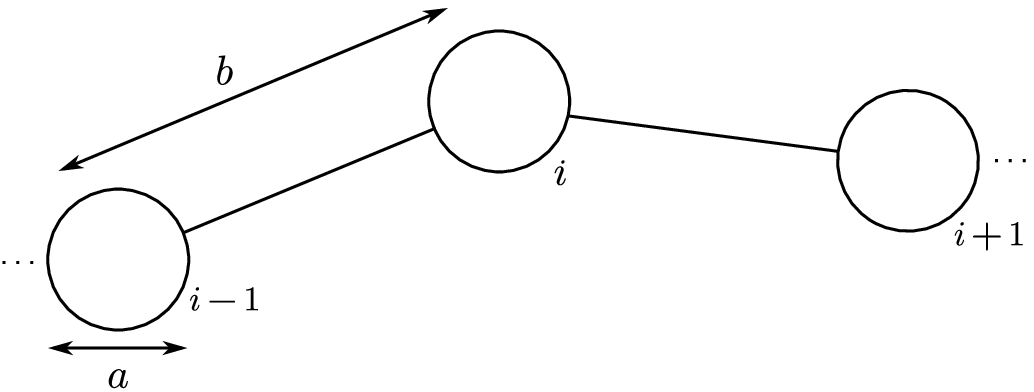}
\put(3,30){{\small (a)}}
\end{overpic}

\begin{overpic}[width=3.25 in]{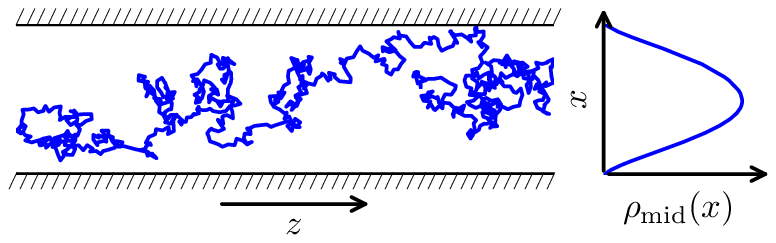}
\put(3,24){{\small (b)}}
\put(92,26){{\small (c)}}
\end{overpic}
 \caption{(a) An illustration of the polymer model, showing the definition of the parameters $a$ and $b$. (b) A snapshot (2D projection) of a configuration resulting from the Monte Carlo program. (c) The probability distribution of the middle monomer, $\rho_{\rm mid}(x)$, as determined from simulations. The parameters for (b) and (c) are $N=800$, $D=22 b$, $a=0.77b$.}
 \label{fig:illustration}
\end{figure}

\noindent In general, the monomer distributions of a polymer confined to a channel will depend on the details of the polymer model, but not if the channels size ($D$) is much larger than the Kuhn length ($b$) of the monomer. In this limit, we can therefore restrict our analysis and simulations to the simple case of a self-avoiding, freely jointed polymer, consisting of $N$ monomers of length $b$. Self-avoidance enters the model as a minimal distance $a$ between the centers of non-neighboring monomers, yielding the excluded volume $\xi=4\pi a^3/3$ for each monomer. This model is illustrated in \subfig{illustration}{a}. By changing $a$, the effect of self-avoidance on the monomer distributions can be modified. We assume that the polymer does not interact with the walls of the channel, except that the center of a monomer is constrained to lie within the interior of the channel. 
Unless otherwise stated, our simulations use the parameters $N=800$ for the number of Kuhn length monomers in a polymer, and $D=22 b$ for the channel dimensions. \subfig{illustration}{b} shows an example of a configuration, for a polymer with $a=0.77b$.\\ \hspace*{\fill}

\begin{figure}[t]
 \begin{overpic}[width=3.25 in]{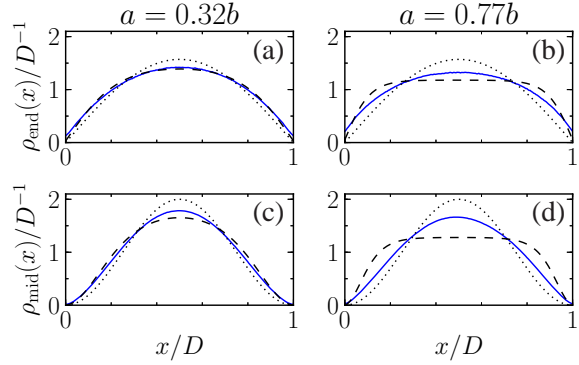}
 \put(43,53){{\small (a)}}
 \put(88,53){{\small (b)}}
 \put(43,27){{\small (c)}}
 \put(88,27){{\small (d)}}
 \end{overpic}
 \caption{The distribution of the end monomer (a, b) and the middle monomer (c, d) of a freely jointed chain of $N=800$ monomers of length $b$ in a channel of diameter $D=22 b$, for two different values of the self-avoidance parameter $a$. (a, c): $a=0.32b$. (b, d): $a=0.77b$. Monte Carlo simulations (solid line) compared to the distribution predicted by the self-consistent field theory (dashed line). For comparison, we also show the predicted distributions for the ideal chain (dotted line).}
 \label{fig:midAndEndDistribution_simulationsAndDeGennes}
\end{figure}

\noindent The cross-sectional distribution of a monomer depends on the position it occupies in the polymer. \fig{midAndEndDistribution_simulationsAndDeGennes} shows the distributions of the end and middle monomer of a self-avoiding chain, for two values of the self-avoidance parameter $a$. We compare the results from the Monte Carlo simulations to the predictions from self-consistent field theory and the ideal diffusion theory, both described below. While the agreement is good for small $a$, self-consistent theory always overestimates the broadening effect of the self-avoidance.
\section{Theoretical predictions for the monomer distributions of a confined polymer}

\noindent The starting point for understanding the monomer distributions of a confined polymer is the propagator $G_{N}(\rv;\rn)$, defined as the combined statistical weight of all polymers with one end at $\rn$ and the other other end at $\rv$\cite{deGennesBook}.
If the propagator is known, the distribution of an end monomer can be obtained by integrating the propagator over all starting positions: $\rho_N(\rv)\propto\int\! d\rn G_{N}(\rv;\rn)$. For monomer $n$ in the interior of the polymer, the distribution can be obtained by treating the polymer as the concatenation of two shorter polymers, yielding $\rho_{n;N}(\rv)\propto \int\! d\rn G_n(\rv;\rn)\int\!d\rp G_{N-n}(\rv;\rp)$. For this expression to be exact, the propagator corresponding to the second polymer should be computed in the presence of the first polymer. In the following, this complication is disregarded. \\ \hspace*{\fill}

\noindent For an ideal chain, the propagator $G_n(\rv;\rn)$ approximately obeys a diffusion equation:
\begin{equation}
\label{eqn:diffusionEquation}
-\partial_n G_{n}(\rv;\rn)=-b^2/6\nabla_\rv^2G_{n}(\rv;\rn),
\end{equation}
with absorbing boundary conditions at the walls\cite{diMarzio1965}.
Here the operator $\nabla_\rv^2$ acts on the first argument of $G$. For a polymer confined to a square channel of side length $D$, \eq{diffusionEquation} has the well-known solution
\begin{align}
 G_{n}(\rv;\rn) &= G_n^\perp(x;x_0)G_n^\perp(y;y_0)G_n^\parallel(z;z_0), \\
 \label{eqn:GnPerpSumOfEigenfunctions} 
 G_n^\perp(x;x_0) &=\sum_{k=1}^ \infty\frac{2}{D}\sin\frac{k\pi x_0}{D}\sin\frac{k\pi x}{D}\exp\left\{-\frac{b^2\pi^2k^2 n}{6D^2}\right\}, \\
 G_n^\parallel(z;z_0)&=\sqrt{\frac{3}{2\pi n b^2}}\exp\left\{-\frac{3(z-z_0)^2}{2 n b^2}\right\}.
\end{align}
If $n>(6D^2)/(b\pi)^2$, \eq{GnPerpSumOfEigenfunctions} is greatly simplified, since the sum is dominated by the term where $k=1$. For a long chain, the distribution of a monomer at the end of the polymer and in the middle of the polymer therefore have the simple forms
\begin{align}
 \rho_{\rm end}(x) &=\frac{\pi}{2 D}\sin\frac{\pi x}{D}, \\
 \rho_{\rm mid}(x) &=\frac{2}{ D}\sin^2\frac{\pi x}{D}.
 \label{eqn:rhoMidIdeal}
\end{align}
These expressions are shown in \fig{midAndEndDistribution_simulationsAndDeGennes}.

It is important to consider under which conditions \eq{diffusionEquation} is a reasonable simplification. For a self-avoiding polymer, \eq{diffusionEquation} can be modified to include an annihilation term, proportional to the probability of collisions \cite{diMarzio1965}:
\begin{equation}
\label{eqn:ReactionDiffusionEquation_pColl}
-\partial_n G_{n}(\rv;\rn)= \left[-b^2/6\nabla_\rv^2+P_n(\rv;\rn)\right]G_{n}(\rv;\rn).
\end{equation}
$P_n(\rv;\rn)$ is referred to as the \textit{collision probability}. It is defined as the probability that a collision would occur if one were to add a randomly oriented monomer at $\rv$ to a polymer of length $n-1$, starting at $\rn$.  \eq{ReactionDiffusionEquation_pColl} shows that \ref{eqn:diffusionEquation}-\ref{eqn:rhoMidIdeal} are expected to work well if $P_n G_n\ll \left|b^2/6 \nabla_\rv^2 G_n\right|\Leftrightarrow P_n\ll \pi^2 b^2/(6 D^2)$. \\ \hspace*{\fill}

\noindent To bring closure to \eq{ReactionDiffusionEquation_pColl} in the case when this condition is not fulfilled, one must find a relation between $G_n(\rv,\rn)$ and $P_n(\rv;\rn)$. The conventional way of achieving this closure is called self-consistent field theory \cite{deGennesBook}. As mentioned above, this theory rests on the mean field assumption that the collision probability at a given position is proportional to the local monomer density. 
As in the case of the ideal polymer, the distribution of monomer $n$ in the interior of the chain is related to the propagator by $\rho_{n;N}(\rv)\propto \int\! d\rn G_n(\rv;\rn)\int\!d\rp G_{N-n}(\rv;\rp)$. Since the propagator $G_n$ approaches an asymptotic shape as $n\to \infty$, the distributions of all monomers far from the ends are expected to be identical. By assuming that the collision probability for a long chain is proportional to this universal distribution $\rho_{\rm mid}$, a non-linear eigenvalue equation for the propagator can be derived and numerically solved (see supporting information).

\section{The collision probability $P_n$}
\begin{figure}[t]
 \begin{overpic}[width=3.25 in]{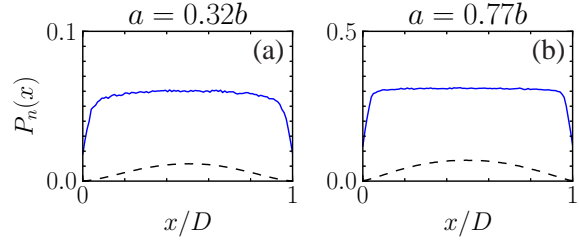}
 \put(43,33){{\small (a)}}
 \put(88,33){{\small (b)}}
 \end{overpic}
 \caption{The collision probability for the end monomer of a freely jointed chain of $N=800$ monomers of length $b$ in a channel of diameter $D=22 b$, for two different values of the self-avoidance parameter $a$. (a): $a=0.32b$. (b): $a=0.77b$. Monte Carlo simulations (solid line) compared to the mean field assumption $P_n=\xi \rho_z \rho_{\rm mid}$ (dashed line).}
 \label{fig:pnEqualsXiRho}
\end{figure}
To test how well the self-consistent assumption for the collision probability works, we performed Monte Carlo simulations for a chain where overlaps between monomers are forbidden, except that the end monomers are allowed to overlap with any other monomer. By measuring the frequency of such allowed overlaps at different positions within the channel, this model allows us to measure the collision probability $P_n(\rv;\rn)$ in simulations. 

In the simulation results presented here, we have only considered the cross-sectional position dependence of the collision probability, which for a long chain is independent of starting position. We denote this quantity $P_n(x,y)$.
\fig{pnEqualsXiRho} shows the marginal collision probability $P_n(x)=\int\! dy P_n(x,y) \rho_{\rm end}(x,y)$. The result is compared to the assumption of the mean-field theory, $P_n(x,y)= \xi \rho_z \rho_{\rm mid}(x,y)$. Here $\rho_z$, the number of monomers per unit channel length, is estimated from simulations as $\rho_z=N/(\mbox{Chain extension})$. \fig{pnEqualsXiRho} shows that the mean-field theory underestimates the collision probability by almost one order of magnitude, yet it overestimates the dependence on position~($x$).

That a mean-field theory for the collision probability cannot be correct is actually apparent from \fig{illustration} -- this snapshot from simulations clearly shows that the average number of other monomers in the vicinity of a given monomer is significantly higher than one would expect from the average concentration $\rho(\rv)$. The reason is clear: since the monomers are connected, if we know that one monomer is located at $\rv$, closely neighboring monomers on the chain will necessarily be close by. Yet this correlation between neighboring monomers is neglected in the mean-field theory.

\begin{figure}[t]
 \begin{overpic}[width=6 in]{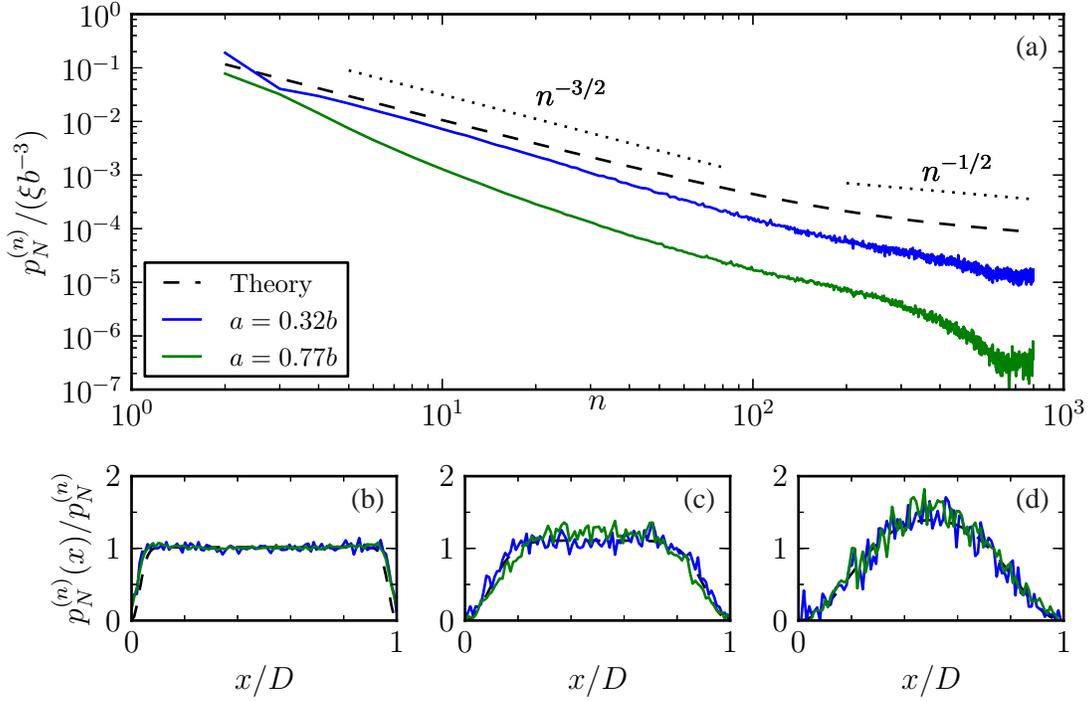}
 \put(91,59){{\small (a)}}
 \put(33,19.5){{\small (b)}}
 \put(62,19.5){{\small (c)}}
 \put(91,19.5){{\small (d)}}
 \end{overpic}
 \caption{The collision probability $\pnn(x,y)$ that an end monomer with $\rv=(x,y,z)$ collides with a monomer $n$ steps distant, for two different values of the self-avoidance parameter $a$. Blue lines: $a=0.32b$. Green lines: $a=0.77b$. Monte Carlo simulations (solid lines) compared to the collision probability predicted by the \eq{pColl_n_idealSum} (dashed lines). (a): $p_N^{(n)}=\int\! dx dy \pnn(x,y)\rho_{\rm end}(x,y)$. Dotted lines show the asymptotic scaling of \eq{pColl_n_idealSum}, for large and small $n$. 
 (b): $p_N^{(n)}(x)/p_N^{(n)}$, for $n=5$. (c): $p_N^{(n)}(x)/p_N^{(n)}$, averaged for $n=45-55$. (d): $p_N^{(n)}(x)/p_N^{(n)}$, averaged for $n=190-210$.}
 \label{fig:pColl_simulationsAndIdealTheory_multiplot}
\end{figure}

The reason that the self-consistent field assumption concerning the relation between the collision probability and the monomer distributions does not hold can thus be understood by considering \textit{which} other monomers the end monomer most often collides with. 
Let us denote by $\pnn(\rv;\rn)$ the probability that the final monomer of a chain of length $N$ collides with a monomer $n$ steps removed along the chain. The total collision probability $P_n$ is given by the sum $P_n(\rv;\rn)=\sum_n \pnn(\rv;\rn)$. As before, our simulations have only measured the cross-sectional position dependence. The $n$-dependence of $\pnn$ is shown in \subfig{pColl_simulationsAndIdealTheory_multiplot}{a}. Clearly, the collision probability is much higher for monomers which are closely neighboring than for distant ones. 
However, the probability of colliding with a neighbor does not vary with $x$ in the same way as the probability to collide with a distant monomer does, as can be seen in \subfig{pColl_simulationsAndIdealTheory_multiplot}{b-d}, where the collision probability $p_N^{(n)}(x)=\int\! dy \pnn(x,y)\rho_{\rm end}(x,y)$ of colliding with a monomer $n$ steps distant from the end monomer is shown, for different values of $a$ and $n$. In particular, for nearby monomers the collision probability is larger close to the walls than the self-consistent field theory predicts. \\\hspace*{\fill}

\noindent The collision probabilities $\pnn(x,y)$ can be qualitatively explained. The starting point is the observation that $\pnn$ must be given by 
\begin{equation}
 \label{eqn:pnnApproxXiRhonn}
 \pnn(\rv;\rn)=\int_\xi\! d\rpr \rhonn(\rv;\rpr;\rn)\approx \xi \rhonn(\rv;\rv;\rn).
\end{equation}
Here $\rhonn(\rv;\rpr;\rn)$ is the probability density that a monomer of length $N$, starting at $\rn$ and ending at $\rv$, has monomer $N-n$ at location $\rpr$.
The integral runs over the excluded volume of the final monomer.
$\rhonn(\rv;\rpr;\rn)$ is related to the propagator of the self-avoiding chain by
\begin{equation}
\label{eqn:rhonnEqualsGGG}
 \rhonn(\rv;\rpr;\rn)=\frac{G_{N-n}(\rpr;\rn)G_n(\rv;\rpr)}{G_N(\rv;\rn)}.
\end{equation}
Unfortunately, we do not know the propagator for the self-avoiding chain. Yet for small values of $\xi$ it may be a good approximation to use the ideal propagator, \eq{GnPerpSumOfEigenfunctions}. Inserting these expressions for the propagator into \ref{eqn:pnnApproxXiRhonn} and \ref{eqn:rhonnEqualsGGG}, and assuming that $N-n\gg\frac{6D^2}{\pi^2b^2}$, we arrive at a collision probability which depends only on the $x$- and $y$-component of $\rv$,
\begin{equation}
\label{eqn:pColl_n_idealSum}
 p_N^{(n)}(x,y)=\xi \sqrt{\frac{3}{2\pi n b^2}}\sum_{k_x=1}^ \infty\sum_{k_y=1}^ \infty\left(\frac{2}{D}\right)^2\sin^2\frac{k_x\pi x}{D}\sin^2\frac{k_y\pi y}{D}\exp\left\{-\frac{b^2\pi^2(k_x^2+k_y^2-2) n}{6D^2}\right\}.
\end{equation}
Here the prefactor $\sqrt{\frac{3}{2\pi n b^2}}=G_n^\parallel(z;z)$ is the return probability of the one dimensional random walk performed by the $z$-component of the ideal chain.
This theory for the collision probability is compared to the results of simulations in \fig{pColl_simulationsAndIdealTheory_multiplot}. \subfig{pColl_simulationsAndIdealTheory_multiplot}{a} shows that the theory can explain qualitatively how the collision probability decreases with contour distance, although \eq{pColl_n_idealSum} underestimates the rate at which $\pnn$ increases with separation ($n$). Even better, it explains quantitatively how the monomer collision probability depends on the position in the channel [\subfig{pColl_simulationsAndIdealTheory_multiplot}{b-d}], showing that the collision probability of nearby monomers is essentially independent of position, except very close to the walls [\subfig{pColl_simulationsAndIdealTheory_multiplot}{b}]. As the contour distance increases, the position dependence increases [\subfig{pColl_simulationsAndIdealTheory_multiplot}{c}]. 
For very large contour separations ($n>6D^2/(\pi b^2)$), the positions are no longer correlated, and the mean field assumption $\pnn(x)\propto \rho_{\rm mid}(x)\approx 2/D\sin^2(\pi x/D)$ works well [\subfig{pColl_simulationsAndIdealTheory_multiplot}{d}]. Since the probability of colliding with neighboring monomers is much higher than the probability of remote collisions, the total collision probability $P_n=\sum_n \pnn$ is dominated by the terms with small values of $n$. Since these terms depend only weakly on position, the same will be true of the total collision probability. This is clearly seen in \fig{pnEqualsXiRho}. \\\hspace*{\fill}

\begin{figure}[t]
 \begin{overpic}[width=3.25 in]{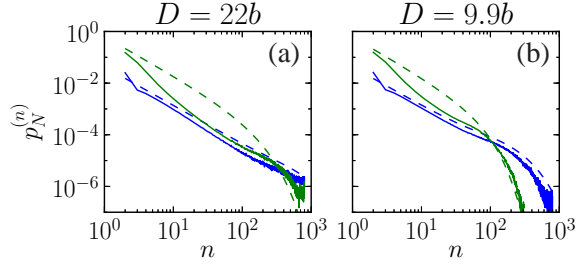}
 \put(46,36){{\small (a)}}
 \put(86,36){{\small (b)}}
 \end{overpic}
 \caption{The collision probability for the end monomer of a freely jointed chain of $N=800$ monomers, in a channel of diameter $D=22 b$ (left) and $D=9.9 b$ (right), for two different values of $a$. (a): $a=0.32b$. (b): $a=0.77b$. Monte Carlo simulations (solid lines) compared to the collision probability predicted by the biased random walk model (dashed lines). Blue lines: $a=0.32 b$. Green lines: $a=0.77 b$. Note that the collision probability is not normalized by $\xi$, as in \fig{pColl_simulationsAndIdealTheory_multiplot}.}
 \label{fig:pColl_simulationsAndBRW}
\end{figure}

\noindent For large separations $n$, \eq{pColl_n_idealSum} predicts that $p_N^{(n)}\propto n^{-1/2}$. Yet this cannot be the whole story, since $\int n^{-1/2} dn$ diverges at infinity. Thus, \eq{pColl_n_idealSum} must break down for $n$ above a certain threshold $n_c$, which we estimate by modeling the diffusion in the $z$-direction as a biased random walk \cite{werner2012}. This model was developed for semi-flexible chains, but it can easily be adapted for weakly self-avoiding freely jointed chains. In this model, each monomer is randomly oriented, except that self-avoidance causes a bias in either the positive or negative $z$-direction, shared by all monomers. The $z$-coordinate thus performs a biased random walk, and the probability density of any subsequence of $n$ monomers starting at $z_0$ and ending at $z$ is given by
\begin{equation}
 \rho_n^\parallel(z;z_0)=\frac{1}{\sqrt{2\pi n\sigma^2}}\exp\left(-\frac{(z-z_0-n\mu)^2}{2 n\sigma^2}\right),
\end{equation}
where $\mu$ is the bias and $\sigma^2$ the variance of each step in the random walk. For broad channels, $D\gg b$, $\sigma^2=b^2/3$, and the bias can be estimated by a mean field argument\cite{werner2012}. The result is $\mu= [{\xi b^2}/{(6D^2)}]^{1/3}$. To compute the collision probability, the relevant quantity is the return probability density
\begin{equation}
\label{eqn:returnProb_BRW}
 \rho_n^\parallel(z;z)=\frac{1}{\sqrt{2\pi n\sigma^2}}\exp\left(-\frac{\mu^2}{2 \sigma^2}n\right)=\sqrt{\frac{3}{2\pi nb^2}}\exp\left(-\frac{n}{n_c}\right),
\end{equation}
where 
\begin{equation}
 \label{eqn:nc}
n_c=\frac{2\sigma^2}{\mu^2}=\frac{2}{3}\left(\frac{6bD^2}{\xi}\right)^{2/3}.
\end{equation}
\eq{returnProb_BRW} agrees with the ideal return probability $\sqrt{\frac{3}{2\pi n b^2}}$ for $n\ll n_c$, but is exponentially suppressed for large values of $n$. In \fig{pColl_simulationsAndBRW}, the measured collision probability is  compared with \eq{pColl_n_idealSum}, modified to include the exponential cutoff. Since the length of the chain is too short to see a clear effect of the cutoff for the large channels we have considered so far [\subfig{pColl_simulationsAndBRW}{a}], we have also performed simulations for identical chains in a narrower channel, where according to \eq{nc} $n_c$ is smaller. For the narrower channels, it is evident that the cutoff is very well explained by the biased random walk model [\subfig{pColl_simulationsAndBRW}{b}]. \\\hspace*{\fill}

\begin{figure}[t]
 \begin{overpic}[width=3.25 in]{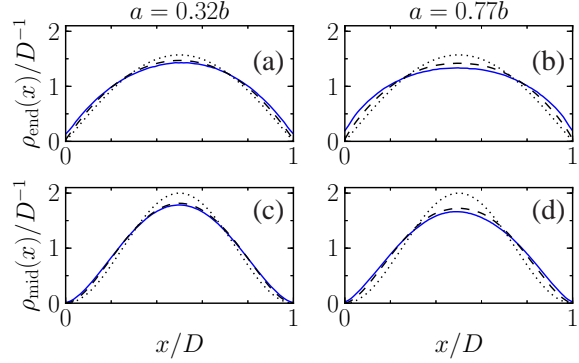}
 \put(43,51){{\small (a)}}
 \put(88,51){{\small (b)}}
 \put(43,27){{\small (c)}}
 \put(88,27){{\small (d)}}
 \end{overpic}
 \caption{The distribution of the end monomer (a, b) and the middle monomer (c, d) of a freely jointed chain of $N=800$ monomers of length $b$ in a channel of diameter $D=22 b$, for two different values of the self-avoidance parameter $a$. (a, c): $a=0.32b$. (b, d): $a=0.77b$. Monte Carlo simulations (solid line) compared to the distribution predicted by the diffusion-annihilation equation with the measured collision probability as the sink term (dashed line). For comparison, we also show the predicted distributions for the ideal chain (dotted line).}
 \label{fig:midAndEndDistribution_compareExactAndSimulations}
\end{figure}

\noindent Whereas the biased random walk model for the collision probability is qualitatively correct, it is not precise enough to quantitatively explain the monomer distributions of the self-avoiding chain. However, we can still pose the question whether the diffusion-annihilation equation correctly describes the distribution of an end monomer, and whether the probability density of a middle monomer is proportional to the square of the probability density of an end monomer. To answer this question, we have numerically solved \eq{ReactionDiffusionEquation_pColl}, using the value of $P_n(x,y)$ measured in our simulations as input. 
The results are seen in \fig{midAndEndDistribution_compareExactAndSimulations}, they agree very well with corresponding results of the simulations, even for the large value of $a$. As in the ideal case, the computed monomer distributions are somewhat too low close to the walls and too high in the center, which again can be explained by the fact that the diffusion approximation fails close to the wall and that both curves are normalized. 
Apart from the good agreement the most striking feature of the monomer distributions is how well the ideal theory works, even for the larger value of $a$. This is surprising, since we saw before that the ideal theory is expected to work well only when $P_n\ll \pi^2 b^2/(6 D^2)$. Yet \fig{pnEqualsXiRho} shows that $P_n \approx 0.05$ for $a=0.32b$ and $P_n \approx 0.3$ for $a=0.77b$ -- more than one order of magnitude \textit{larger} than $\pi^2 b^2/(6 D^2)\approx 0.003$. The resolution of this apparent paradox is that although $P_n(x)$ is large, it is almost constant except very close to the walls (because most collisions are with nearby monomers, for which $\pcn(x)$ depends only weakly on $n$). 
Yet adding a constant term to $P_n(x,y)$ only changes the normalization of the solution to \eq{ReactionDiffusionEquation_pColl}, but does not change its shape (assuming $P_n\ll 1$ still holds). Since changing the normalization of $G_n$ has no impact on the monomer distributions, neither has the constant term of the collision probability. Since the collision probability is almost constant in the interior of the channel (\fig{pnEqualsXiRho}), it will influence the monomer distributions only weakly. It is true that the sharp decrease of the collision probability close too the walls will cause the \textit{relative} monomer density to increase significantly there, but this increase is very small in absolute terms.
Thus, the effect of self-avoidance upon the monomer distributions is much smaller than one might expect. 

\section{Discussion}
\noindent In this paper we have measured through Monte Carlo simulations the monomer distributions of a long, self-avoiding polymer confined to a square channel of size $D\gg b$, where $b$ is the Kuhn length of the polymer. We find that the distributions are qualitatively similar to the distributions of an ideal chain, which can be determined from the solution of a diffusion equation with absorbing boundary conditions\cite{casassa1967}. This similarity holds even for strongly self-avoiding polymers, for which a priori the ideal approximation would be expected to fail. In fact, the distributions measured in simulations are more similar to the prediction of the ideal theory, than to the predictions of self-consistent field theory, which includes the effect of self-avoidance on the monomer distributions by means of a mean-field theory for the probability of intra-chain collisions\cite{deGennesBook}.

To explain this surprising finding, we have analyzed how the probability of collisions between two monomers depends on how close they are located along the polymer, and upon their position in the channel. We have found that this collision probability can be qualitatively explained by calculating the return probability of a particle diffusing with a drift in the channel direction. Our results show that most collisions involve closely neighboring monomers, for which the collision probability cannot be described accurately by the mean-field theory. For these monomers, the collisions probability is much higher than the mean-field theory predicts, but depends only weakly on the position in the channel. We show that this weak dependence on position explains why self-avoidance has such a small impact on the monomer distributions.

We conclude with a number of comments and questions. First, we have shown that given the measured collision probability $P_n(x,y)$, the distribution of the end monomer of a freely jointed chain solves the diffusion-annihilation equation [\eq{ReactionDiffusionEquation_pColl}]. Yet to bring closure to this problem one must also determine how to compute the collision probability from the propagator $G_n$, we do not know how to do this in a quantitative fashion.

Second, real polymers also involve interactions between neighboring monomers. In particular, real polymers are usually stiffer than a freely jointed chain. In this case, the diffusion theory must be phrased in terms of \textit{effective monomers}, each with a length corresponding to the Kuhn length. A case of particular interest is that of a worm-like chain with persistence length $\lp$ and effective width $\we$, a common model of a DNA molecule in solution. In this case, the effective monomers have a Kuhn length of $2\lp$, and the excluded volume of an effective monomer can be approximated by that of a cylinder, $\xi= \pi w_{\rm eff}(2\lp^2+\frac{1}{2}(\pi+3)\lp w_{\rm eff} + \frac{1}{8}\pi w_{\rm eff}^2)$ \cite{onsager1949}. The results presented for $a/b=0.32$ and $a/b=0.77$ would then correspond to a worm-like chain with $\we/\lp=0.1$ and $\we/\lp=0.9$, respectively.  
The former value is typical for double-stranded DNA experiments performed at high ionic strength\cite{zhang2008, reisner2005}, whereas the latter corresponds to a very flexible polymer, such as single-stranded DNA\cite{tinland1997}.

Third, just as the interaction between different monomers usually is more complicated than hard-core repulsion, so might the interaction between each monomer and the channel walls be. If the interaction is short-ranged and repulsive, the analysis presented here is expected to work if the channel diameter $D$ is replaced by an \textit{effective diameter} $D_{\rm eff}=\int_0^D \! dx \exp(-U(x)/kT)$, where $U(x)$ is the interaction energy between the walls and a monomer at position $x$.

Fourth, we have seen that the self-consistent field theory cannot explain the monomer distributions of a single polymer confined to a channel (except in the limit of extremely weak self-avoidance, for which the ideal theory already works very well). Could it work for other geometries?
In order for the monomer distributions to obey self-consistent field theory, most collisions must occur with distant monomers. For a polymer near a wall this is impossible, since the collision probability decreases faster than $n^{-3/2}$, the integral of which does not converge. 
For a chain confined to a slit, the overlap probability of an ideal chain is marginally divergent, $p\propto n^{-1}$. Therefore, any perturbation makes the integral over collision probabilities convergent, and long-range collisions can therefore only dominate the collision probability if the excluded volume is very small. For a chain that is constrained in two dimensions (a channel), we have seen that while there is a theoretical limit in which the self-consistent theory is valid, it is of no practical significance. 

It remains to consider three-dimensional confinement, e.g. a chain confined to a sphere or a cube. In this case, most collisions of a long chain can occur between distant monomers, and the mean-field theory might be expected to work. However, a complicating factor is that unless $\xi$ is small, a long chain will fill the available volume densely, whereas the theory is only valid in the dilute limit ($P_n\ll 1$).

Fifth, what changes if instead of a single polymer, we consider a dilute blend of polymers in a channel? The end monomer can now also collide with monomers from other chains, and its distribution must change as a result. However, since self-avoiding polymers in a channel tend to separate\cite{jun2006}, the presence of other monomers cannot influence the behavior of an interior monomer (unless the polymer is significantly compressed by the other polymers). Thus, the distributions of most monomers will be unchanged by the addition of other polymers. 

Sixth, we have seen that a self-avoiding chain in a channel can be modeled as a biased random walk. The strength of the bias is related to the probability of collisions, and can be estimated by a mean field argument \cite{werner2012}. However, the mean field argument assumes that monomers are uniformly distributed in the channel, and should therefore underestimate the collision frequency, and hence the bias. Also, since it is a mean-field theory, it does not necessarily describe the fluctuations or the distribution of the end-to-end distance correctly. Therefore, the exact shape of the cutoff in $\pcn$ is not necessarily exponential, and the cutoff distance $n_c$ need not be related to the bias exactly as in \eq{nc}. 
Another problem is that recent simulations\cite{werner2012} show that whereas the biased random walk model works well for the monomers in the interior of the chain, monomers close to the end show a smaller bias. This fact should modify the probability of colliding with distant monomers, but the size of this effect is hard to estimate. The source of this \textit{end decay} of the bias is not known, but might well be related to the fact that end monomers are distributed differently than interior monomers, as shown in \fig{midAndEndDistribution_compareExactAndSimulations}.

Seventh, we have, throughout this paper, assumed that $D\gg b$, so that the diffusion approximation can be applied. It would, however, be interesting to consider the question of how the theory presented here fails outside this asymptotic limit. This would depend sensitively on the chain model, the interactions between monomers, and between the polymer and the walls of the channel. A particularly interesting case is that of a DNA molecule confined  in a nanochannel. Such a molecule is usually modeled as a worm-like chain of persistence length $\lp\approx \SI{50}{\nano\meter}$, with screened repulsive electrostatic interactions between distant monomers, and between the molecule and the walls. The question of how such a molecule behaves when confined in channel sizes of the same order as the persistence length is of great experimental and biological interest. The resolution in standard fluorescence microscopy is limited by the diffraction of light, which limits the possibility to experimentally observe these 
distributions.
However, recent experimental developments make it possible to directly observe DNA molecules with a resolution below the diffraction limit of light \cite{persson2011}, 
and it should therefore be possible to test present and future theoretical predictions for the monomer distributions against experimental data.

\section{Acknowledgments}
This work was supported by Vetenskapsr\aa{}det, the Chalmers Area of Advance in Nanoscience and Nanotechnology, and the G\"oran Gustafsson Foundation for Research in
Natural Sciences and Medicine.
\\\hspace*{\fill}

\noindent \textbf{Supporting Information Available}: Details of the Monte Carlo program and the numerical solution to the self-consistent field equation are given as electronic supporting information. This material is appended at the end of the document.

\providecommand{\refin}[1]{\\ \textbf{Referenced in:} #1}

\end{document}


\date{\today}
\section{Electronic supporting information}
\subsection{Numerical solution of the self-consistent field equation}
Assuming that the density is constant along the $z$-direction, the self-consistent field equation\cite{deGennesBook} for a polymer confined to a square channel is
\begin{equation}
\label{eqn:selfConsistentFieldEquation}
 \lambda u(x,y)=\left[-\frac{b^2}{6}\left(\frac{\partial^2}{\partial x^2}+\frac{\partial^2}{\partial y^2}\right)+\xi\rho_z u^2(x,y)\right]u(x,y),
\end{equation}
with boundary conditions
\begin{equation}
u(x,0)=u(x,D)=u(0,y)=u(D,y)=0.
\end{equation}
Here, the eigenvalue $\lambda$ is a free parameter, the parameters $b$ and $\xi$ are given by the polymer model, and $\rho_z$ is estimated from simulations as $\rho_z=N/(\mbox{Chain extension})$. Since this eigenvalue equation is non-linear, it is important to note that the eigenfunction $u(x,y)$ must be normalized, 
\begin{equation}
 \iint_0^D\! dx dy [u(x,y)]^2 = 1.
\end{equation}\hspace*{\fill}

\noindent We have solved this equation numerically by an iterative procedure. Starting with $u_0(x,y)=(2/D) \sin(\pi x/D)\sin(\pi y/D)$ (the ground state of the diffusion equation), we updated the solution according to
\begin{equation}
 u_{n+1}=\frac{1}{Z_{n+1}}\left[(1-\alpha)u_n+\alpha u^\prime_n\right],
\end{equation}
where $z_{n+1}$ is a normalization constant. $u^\prime_n$ is the normalized solution to the (linear) eigenvalue equation
\begin{equation}
\label{eqn:linearEigenvalueEquation}
 \lambda u^\prime_n(x,y)=\left[-\frac{b^2}{6}\left(\frac{\partial^2}{\partial x^2}+\frac{\partial^2}{\partial y^2}\right)+\xi\rho_z \left[u_n(x,y)\right]^2\right]u^\prime_n(x,y).
\end{equation}
The convergence constant $\alpha$ is necessary for the algorithm to converge to a solution. For the parameters we have considered, $\alpha=0.5$ is enough to ensure convergence.

\eq{linearEigenvalueEquation} was solved by discretizing the region $0<x,y<D/2$ into a grid of $100\times100$ squares and representing the function $u^\prime_n(x,y)$ as a vector of length $100\times 100=10000$. Treating this region suffices, since the other three quarters are determined by symmetry. In this representation, the operator
\begin{equation}
 \left[-\frac{b^2}{6}\left(\frac{\partial^2}{\partial x^2}+\frac{\partial^2}{\partial y^2}\right)+\xi\rho_z \left[u_n(x,y)\right]^2\right],
\end{equation}
with appropriate boundary conditions, is a sparse matrix of size $10000\times10000$. The eigenvector corresponding to its smallest eigenvalue gives the solution $u^\prime_n(x,y)$. This eigenvector was determined by using the subroutine \textit{eigsh} of Numerical Python (NumPy)\cite{scipy}.
\subsection{Details of the Monte Carlo program}
Our Monte Carlo simulations implement the Metropolis algorithm, with crankshaft trial updates \cite{yoon1995}. In each update step, $n$ adjacent molecules $i=n_{\rm min},n_{\rm min},\dots, n_{\rm max}=n_{\rm min}+n-1$ are selected, where $1\le n\le 16$. If the $n$ monomers are all in the interior of the chain, they are rotated by a random angle $\theta\in [0,2\pi)$ around the axis connecting monomers $n_{\rm min}-1$ and $n_{\rm max}+1$. If instead $n_{\rm min}=1$ (or $n_{\rm max}=N$) they are rotated by a random angle $\theta\in [0,2\pi)$ around a random axis passing through $n_{\rm max}+1$ (or $n_{\rm min}-1$). 
Note that these trial updates conserve the distance between adjacent monomers. For these simulations, the chain is initialized such that the distance between the center of adjacent monomers is $b$. In order for the chain to equilibrate, $2\cdot 10^{10}$ trial updates are suggested before starting the measurement process. The measurement process runs for $10^{11}$ trial updates, with measurements taken once every $10^4$ steps.

\providecommand*\mcitethebibliography{\thebibliography}
\csname @ifundefined\endcsname{endmcitethebibliography}
  {\let\endmcitethebibliography\endthebibliography}{}